%% file: WZResPaper.tex
\documentclass[aps,prd,twocolumn,showpacs,superscriptaddress,groupedaddress]{revtex4}

\usepackage{graphicx}  
\usepackage{dcolumn}  
\usepackage{bm}       
\usepackage{amssymb} 
\usepackage{multirow}
\usepackage{epsfig}
\usepackage{epstopdf}
\usepackage{amsmath,booktabs,array}
\usepackage[hyperindex]{hyperref} 
\usepackage{atlasphysics}
\usepackage{eqparbox}
\usepackage{subfigure}

\newcommand{\metpaul}{\mbox{$\rlap{\kern0.15em/}E_T$}}

\newcommand{\WZ}{$WZ$}
\newcommand{\alpgen}{{\sc alpgen}}

\newcommand{\jimmy}{{\sc jimmy}}
\newcommand{\mcatnlo}{{\sc mc@nlo}}

\newcommand{\pythia}{{\sc pythia}}

\newcommand{\herwig}{{\sc herwig}}
\newcommand{\sherpa}{{\sc sherpa}}
\newcommand{\geant}{{\sc geant4}}
\newcommand {\Et}   {\ensuremath{E_{\mathrm{T}}}}

\newcommand {\rhot}{\ensuremath{\rho_{\rm T}}}
\newcommand {\Mrho}{\ensuremath{m_{\rho_{\rm T}}}}

\newcommand {\Ma}{\ensuremath{m_{a_{\rm T}}}}
\newcommand {\at}{\ensuremath{a_{\rm T}}}
\newcommand {\mtwz}{\ensuremath{m_{\rm T}^{WZ}}}

\usepackage{preprintcover} 
\PreprintCoverPaperTitle{Search for resonant $WZ$ production in the $WZ
  \rightarrow \ell\nu \ell^{\prime}\ell^{\prime}$ channel in $\sqrt{s}=7$ TeV $pp$ collisions with the ATLAS detector} 

\PreprintIdNumber{CERN-PH-EP-2012-063}  

\PreprintCoverAbstract{A generic search is presented for a heavy particle decaying to 
$WZ \to \ell \nu \ell^{\prime} \ell^{\prime}$ ($\ell, \ell^{\prime} = e, \mu$) 
final states. The data were recorded by the ATLAS detector in 
$\sqrt{s}=7\TeV$ $pp$ collisions at the Large Hadron Collider 
and correspond to an integrated luminosity of 1.02~\ifb. 
The transverse mass distribution of the selected $WZ$ candidates is found to be consistent with
the Standard Model expectation. Upper limits on
the production cross section times branching ratio are derived using two benchmark
models predicting a heavy particle decaying to a $WZ$ pair.} 

\PreprintJournalName{Physical Review D}  

\begin{document}

\title{Search for resonant $WZ$ production in the $WZ
 \rightarrow \ell\nu \ell^{\prime}\ell^{\prime}$ channel in $\sqrt{s}=7$ TeV \\ $pp$ collisions with the ATLAS detector}

\author{The ATLAS Collaboration}

\date{\today}

\begin{abstract}
A generic search is presented for a heavy particle decaying to 
$WZ \to \ell \nu \ell^{\prime} \ell^{\prime}$ ($\ell, \ell^{\prime} = e, \mu$) 
final states. The data were recorded by the ATLAS detector in 
$\sqrt{s}=7\TeV$ $pp$ collisions at the Large Hadron Collider 
and correspond to an integrated luminosity of 1.02~\ifb. 
The transverse mass distribution of the selected $WZ$ candidates is found to be consistent with
the Standard Model expectation. Upper limits on
the production cross section times branching ratio are derived using two benchmark
models predicting a heavy particle decaying to a $WZ$ pair.
\end{abstract}

\maketitle

\section{Introduction}
The study of electroweak boson pair production is a powerful test 
of the spontaneously broken gauge symmetry of the Standard Model (SM) 
and can be used as a probe for new phenomena beyond the SM.  
Heavy particles that can decay to gauge boson pairs are predicted by 
many scenarios of new physics, including the Extended
Gauge Model (EGM)~\cite{Altarelli:1989ff}, Extra
Dimensions~\cite{Randall:1999ee,Davoudiasl:1999tf}, and Technicolor
 models~\cite{Eichten:2007sx,Catterall:2011zf,Andersen:2011yj}. 

This paper describes the search for a resonant structure in  
$WZ \to \ell \nu \ell^{\prime} \ell^{\prime}$ ($\ell, \ell^{\prime}=
e, \mu$) production above 200 \GeV. 
The dataset used corresponds to an integrated luminosity of 1.02 \ifb, collected by the 
ATLAS detector at the Large Hadron Collider in $pp$ collisions at a center-of-mass
energy of $\sqrt{s} = 7$~\TeV\ during the 2011 data taking. 
Events are selected with three charged leptons (electrons or muons) and large missing transverse momentum (\met) 
due to the presence of a neutrino in the final state. 
Two benchmark models, which predict the existence of narrow heavy
particles decaying into $WZ$, are used to interpret the
results: the EGM, through heavy vector boson 
$W'$ production, and the Low Scale Technicolor model
(LSTC)~\cite{Eichten:2007sx}, through technimeson production. 

The couplings of the EGM $W'$ boson to the SM particles are the same as those of the
$W$ boson, except for the coupling to $WZ$, whose strength is $g_{W'WZ} =
g_{WWZ} \times m_W m_Z/m_{W'}^2$, where $g_{WWZ}$ is the SM $WWZ$
coupling strength, and $m_W, m_Z$ and $m_{W'}$ are the masses
of the $W$, $Z$ and $W'$ particles, respectively. 
Strong bounds exist on $m_{W'}$ from $W' \rightarrow \ell \nu$ 
searches~\cite{Aad:2011yg, cms_Wprime, cdf_wprime, d0_wprime} assuming the 
Sequential Standard Model (SSM) as the benchmark model, 
in which the $W'$ coupling to $WZ$ is strongly suppressed. 
The $W' \rightarrow WZ$ search presented in this paper is thus
independent of, and complementary to, $W' \rightarrow \ell\nu$ searches. 
Searches for the EGM $W'$ boson in the $WZ$ channel have been
performed at the Tevatron and $W'$ bosons with a mass between
$180$~\GeV\ and $690$~\GeV\  
are excluded at 95\% confidence level (CL)~\cite{Abazov:2010dj,Aaltonen:2010ws}.

In the LSTC model, technimesons with narrow widths are predicted which
decay to $WZ$. Examples are the lightest vector technirho \rhot\ and 
its axial-vector partner techni-$a$ $a_{\rm T}$. 
A previous search in the $WZ$ decay channel has been performed by the
D0~experiment and \rhot\ technimesons with a mass between $208$~\GeV\ and $408$~\GeV\ are excluded 
at 95\% CL under the specific mass hierarchy assumption $m_{\rho_{\rm
    T}} < m_{\pi_{\rm T}} + m_W$, where $m_{\rho_{\rm T}}$, $m_{\pi_{\rm T}}$ are the masses
of the technirho and technipion, respectively~\cite{Abazov:2009eu}.

\section{The ATLAS Detector}
The \mbox{ATLAS} detector~\cite{atlas_detector} is a general-purpose
particle detector with an approximately forward-backward symmetric 
cylindrical geometry, and almost $4\pi$ coverage in solid angle~\cite{geometry}.
The inner tracking detector (ID) covers the pseudorapidity range of $|\eta|<2.5$ and consists of a silicon
pixel detector, a silicon microstrip detector, and a transition 
radiation tracker. The ID is surrounded by a thin superconducting solenoid providing a
$2 \ $T magnetic field, and by a calorimeter system covering an $\eta$
range up to $4.9$, which provides three-dimensional reconstruction of particle showers. 
For $|\eta|<2.5$, the electromagnetic calorimeter is finely
segmented and uses lead as absorber and liquid argon (LAr) as active material.
The hadronic calorimeter uses steel and scintillating tiles in the barrel region, while 
the endcaps use LAr as the active material and copper as absorber. The forward calorimeter 
also uses LAr as active medium with copper and tungsten as absorber. 
The muon spectrometer (MS) is based on one barrel and two endcap air-core toroids, 
each consisting of eight superconducting coils arranged
symmetrically in azimuth, and surrounding the calorimeter. 
Three layers of precision tracking stations, consisting of drift tubes
and cathode strip chambers, allow a precise muon momentum measurement up
to $|\eta|<2.7$. Resistive plate and thin-gap chambers provide muon triggering capability up to $|\eta|<2.4$.

\section{Monte Carlo Simulation}
\label{sec:mc} 
Monte Carlo (MC) simulated samples are used to model signal and background processes. 
Events are generated at $\sqrt{s} = 7$ \TeV\ and the detector response
simulation~\cite{atlas-sim} is based on the \geant~program~\cite{geant}. 

The simulation of the signals, both for the EGM $W'$ and the LSTC \rhot\ production,
is based on the leading-order (LO) \pythia~\cite{pythia} event generator, with modified 
leading-order (LO$^*$)~\cite{Sherstnev:2008dm} parton distribution function (PDF) set 
MRST2007 LO$^*$~\cite{Thorne:2009ky}. By default, \pythia\ also includes
\at\ production, as discussed below. A mass-dependent $k$-factor is
used to rescale the LO$^*$ \pythia\ prediction to the 
next-to-next-to-leading order (NNLO) cross section. The $k$-factor is computed using
the {\sc zwprod} program~\cite{zwprod} in the approximation of zero-width for the resonance; 
its value decreases with the resonance mass from 1.17 at $m_{W'} = 200 \GeV$ to 1.08 at $m_{W'} = 1 \TeV$. 

The LSTC simulated samples correspond to the following set of parameters: number of technicolors
$N_{TC} = 4$, charges of up-type and down-type technifermions $Q_U=1, Q_D=0$,
mixing angle between technipions and electroweak gauge boson
longitudinal component $\sin\chi=1/3$. 
The \rhot\ can decay both to $WZ$ and $\pi_{\rm T} W$; if the \rhot\
and $\pi_{\rm T}$ masses are degenerate, the branching ratio $BR(\rhot\ \to WZ)$ is 100\%.  
Two-dimensional exclusion regions 
are set on the technicolor production in the ($m_{\rhot}$ ,
$m_{\pi_T}$) plane.  In addition, for comparison purpose with
previous results~\cite{Abazov:2009eu}, the relation $\Mrho = m_{\pi_{\rm T}} + m_{W}$ is used when extracting 
one-dimensional limits on the \rhot\ mass, which entails a value of
$BR(\rhot\ \to WZ) = 98\%$. The axial-vector partner of the \rhot,
the \at, also decays to $WZ$, and depending on its mass, contributes
to the $WZ$ production cross section. 
Two scenarios for the value of the mass of the $a_{\rm T}$ technimeson are considered: 
$m_{a_{\rm T}} = 1.1 \times m_{\rho_{\rm T}}$, which is the standard
value implemented in \pythia,  and $m_{a_{\rm T}} \gg m_{\rho_{\rm
    T}}$, which is simulated by removing the \at\ contribution at the
generator level.

The SM \WZ\ production, which is an irreducible background for this
search, is modeled by the \mcatnlo~event generator~\cite{mcnlo},
which incorporates the next-to-leading-order (NLO) matrix elements into the parton
shower by interfacing to the \herwig~program~\cite{herwig}. 
The underlying event is modeled with \jimmy~\cite{jimmy}. Other SM processes that can mimic 
the same final state include: $ZZ\rightarrow
\ell\ell\ell^{\prime}\ell^{\prime}$, where one of the leptons is not
detected or fails the selection requirements; $Z (\rightarrow \ell\ell)+\gamma$, where the photon 
is misidentified as an electron; and processes with two identified
leptons and jets, namely $Z$ production in
association with jets ($Z+$jets), $t\bar{t}$ and single top events,
where leptons are present from $b$- or $c$-hadron decays or one jet is misidentified as a lepton.
SM $ZZ$ events are simulated at LO using \herwig~and $W/Z+\gamma$
production is modeled with \sherpa~\cite{sherpa}. The cross sections for these 
two processes are corrected to the NLO calculation computed with {\sc mcfm}~\cite{Campbell:1999ah,Campbell:2011bn}. 
The $W/Z+$jets process is modeled at LO using \alpgen~\cite{alpgen}, and then corrected to the NNLO cross section 
computed with {\sc fewz}~\cite{Gavin:2010az}. Single top and $t\bar t$ 
events are simulated at NLO using \mcatnlo.
The backgrounds due to the $Z$+jets, $t\bar{t}$ and single top processes (called 
the ``$\ell\ell^{\prime}+$jets" background in this paper) are 
estimated using data-driven methods and the corresponding MC samples mentioned above are 
used only for cross-checks.

\section{Event Selection}
\label{sec:event_selections} 
The data analyzed are required to have been selected online by a single-lepton ($e$ or $\mu$) trigger with 
a threshold of $20$~\GeV\ on the transverse energy (\Et) in the electron case and $18$~\GeV\ on the transverse
momentum (\pt) in the muon case. 
After applying data quality requirements, the total integrated luminosity of the 
dataset used in this analysis is $1.02 \pm 0.04$ fb$^{-1}$~\cite{Aad:2011dr,ATLAS-CONF-2011-116}. 

Due to the presence of multiple collisions in a single
bunch-crossing, about 6 on average, each event can have
multiple reconstructed primary vertices. The vertex having the largest sum of squared transverse
momenta of associated tracks is selected as the primary vertex of the hard collision and it 
is used to compute any reconstructed quantity referred to the primary interaction vertex. 
To reduce the contamination due to cosmic rays, only
events where the primary vertex of the hard collision has at least three associated tracks
with $\pt > 0.5$~\GeV\ are considered. 
 
Electrons are reconstructed from a combination of an ID
track and a calorimeter energy cluster, with $\Et > 25 \GeV$ and
$|\eta| < 1.37$ or $1.52 < |\eta| < 2.47$, avoiding the transition
region between the barrel and the endcap electromagnetic
calorimeters. Candidate electrons must satisfy the
{\it medium}~\cite{electron_performance} quality definition,  
which is based on the calorimeter shower shape, 
track quality, and track matching with the calorimeter cluster. 
To make sure candidate electrons originate from 
primary interaction vertex, they are also required to have a longitudinal impact
parameter ($|z_0|$) smaller than 10 mm and a
transverse impact parameter ($d_0$) with significance 
($|d_0|/ \sigma_{d_0}$) smaller
than 10, both with respect to the selected primary vertex. 
In addition, the electron is required to be isolated in the calorimeter
such that the sum of the \Et\ of the clusters around the electron within a cone of $\Delta R =
\sqrt{\Delta\eta^2 + \Delta\phi^2}= 0.3$ is less than $4$~\GeV.
Corrections are applied to account for the energy deposition 
inside the isolation cone due to electron energy leakage and additional pile-up collisions.

Muon candidates must be reconstructed in both the ID and the MS, and
the combined track is required to have $\pt > 25$~\GeV\ and $|\eta|<2.4$. 
Good quality is ensured by requiring a minimum number of silicon strip and
pixel hits associated to the track. To suppress the contribution of muons coming from hadronic
jets, the \pt\ sum of other tracks with $\pt > 1 $~\GeV, within a cone of $\Delta R = 0.2$ around the
muon track, is required to be less than 10\% of the muon \pt.
The muon candidate is required to be compatible
with the selected primary vertex, with $|z_0| < 10$ mm and $|d_0| / \sigma_{d_0} <10$. 
 
The missing transverse momentum, \met, is reconstructed, in the range
$|\eta| < 4.5$, as the negative vector sum of calorimeter cell
transverse energies, calibrated to the electromagnetic
scale~\cite{met}, to which the transverse momenta of identified muons
are added.

The $WZ \rightarrow \ell \nu \ell^{\prime} \ell^{\prime}$ candidate events are 
selected by requiring two oppositely-charged same-flavor leptons with an invariant 
mass within $20$~\GeV\ of the $Z$ boson mass, plus a third lepton and $\met > 25$~\GeV. 
The transverse mass of the reconstructed $W$ boson, i.e. $m_{\rm T}^W = \sqrt{2p_{\rm T}^{\ell}\met(1-\cos \Delta \phi)}$, where $p_{\rm T}^{\ell}$ is 
the transverse momentum of the charged lepton and $\Delta \phi$ 
is the opening angle between the lepton and the \met\ direction in the plane
transverse to the beam, is required to be greater than $15$~\GeV\ to
suppress multijet background. Selected events are also required to have 
exactly three charged leptons to suppress the 
$ZZ \rightarrow \ell \ell \ell^{\prime} \ell^{\prime}$ background. 
These selection criteria define the signal region. 
Four decay channels $e\nu ee$, $e\nu\mu\mu$, $\mu\nu ee$ and
$\mu\nu\mu\mu$ are analyzed separately and then combined.
The measurement of the inclusive $pp \rightarrow WZ \rightarrow
\ell\nu\ell^{\prime}\ell^{\prime}$ cross section has previously been reported by ATLAS~\cite{SMWZ}.
This analysis goes further by using the reconstructed event properties to probe for new phenomena.

After the final selection, the transverse mass of the $WZ$ candidates 
(\mtwz) is examined for any resonant structure. 
Here \mtwz~is calculated as $\mtwz = \sqrt{ (E_{\rm T}^{Z} + E_{\rm
  T}^{W})^2 - (p_x^{Z} + p_x^{W})^2 - (p_y^{Z} +
p_y^{W})^2}$, where $E_{\rm T}^{Z}$ and $E_{\rm T}^{W}$ are the  scalar sums of the transverse
energies of the decay products of the $Z$ and $W$ candidates, respectively.
The \met\ vector is used as the estimator of the transverse
momentum of the neutrino arising from the $W$ boson decay.

\section{Background Estimation}
\label{sec:background}

The dominant background for the $WZ$ resonance search comes from SM \WZ\
production. Its contribution is estimated using MC simulation. 
Simulated events are required to pass the event selection criteria and the
final yield is normalized to the integrated luminosity. 
Lepton reconstruction and identification efficiencies, energy scale and 
resolution in the MC simulation are corrected to the corresponding values measured 
in the data in order to improve the overall modeling. 
Other diboson processes such as $ZZ$ and $Z\gamma$ are also estimated 
using MC simulation. 

A data-driven approach is used to estimate the contribution of the 
$\ell\ell^{\prime}+$jets background in the signal region.
It is estimated by selecting a data sample containing two leptons that pass all the 
quality criteria requested in the lepton selection, and a lepton-like
jet, which is defined as a reconstructed object that satisfies all 
quality criteria but fails the electron {\it medium} quality or the
muon isolation requirement. The overall contribution is obtained by scaling each event by a correction factor $f$. 
The factor $f$ is the ratio of the probability for a jet to satisfy the full lepton identification 
criteria to the probability to satisfy the lepton-like jet criteria. 
The factor $f$ is measured both for muons and electrons in a
dijet-enriched data sample as a function of the lepton \pt, and corrected for the 
small contribution of leptons coming from $W$ and $Z$ bosons decays using MC simulation. 

Data and SM predictions are compared in two dedicated signal-free control
regions, selected by requiring the same selection criteria as used for
the signal region except requiring $\mtwz < 300$ GeV for the ``SM $WZ$ control region", 
and requiring $\met<25$~GeV for the ``$\ell\ell^{\prime}+$jets control region". 
The SM $WZ$ control region is used to test 
the modeling of the irreducible background from 
non-resonant $WZ$ production, and the $\ell\ell^{\prime}+$jets
control region is used to assess the 
modeling of the $\ell \ell^{\prime}+$jets background.
Good agreement between data and SM predictions is found in both
control regions, as
shown by the transverse mass distribution of the $W$ boson in the SM $WZ$ control region 
and by the invariant mass distribution of the two leptons coming from the $Z$ boson decay 
in the $\ell\ell^{\prime}+$jets control region displayed in Fig.~\ref{fig:control}.

\begin{figure*}
  \centering
  \subfigure[]{\includegraphics[width=0.49\textwidth]{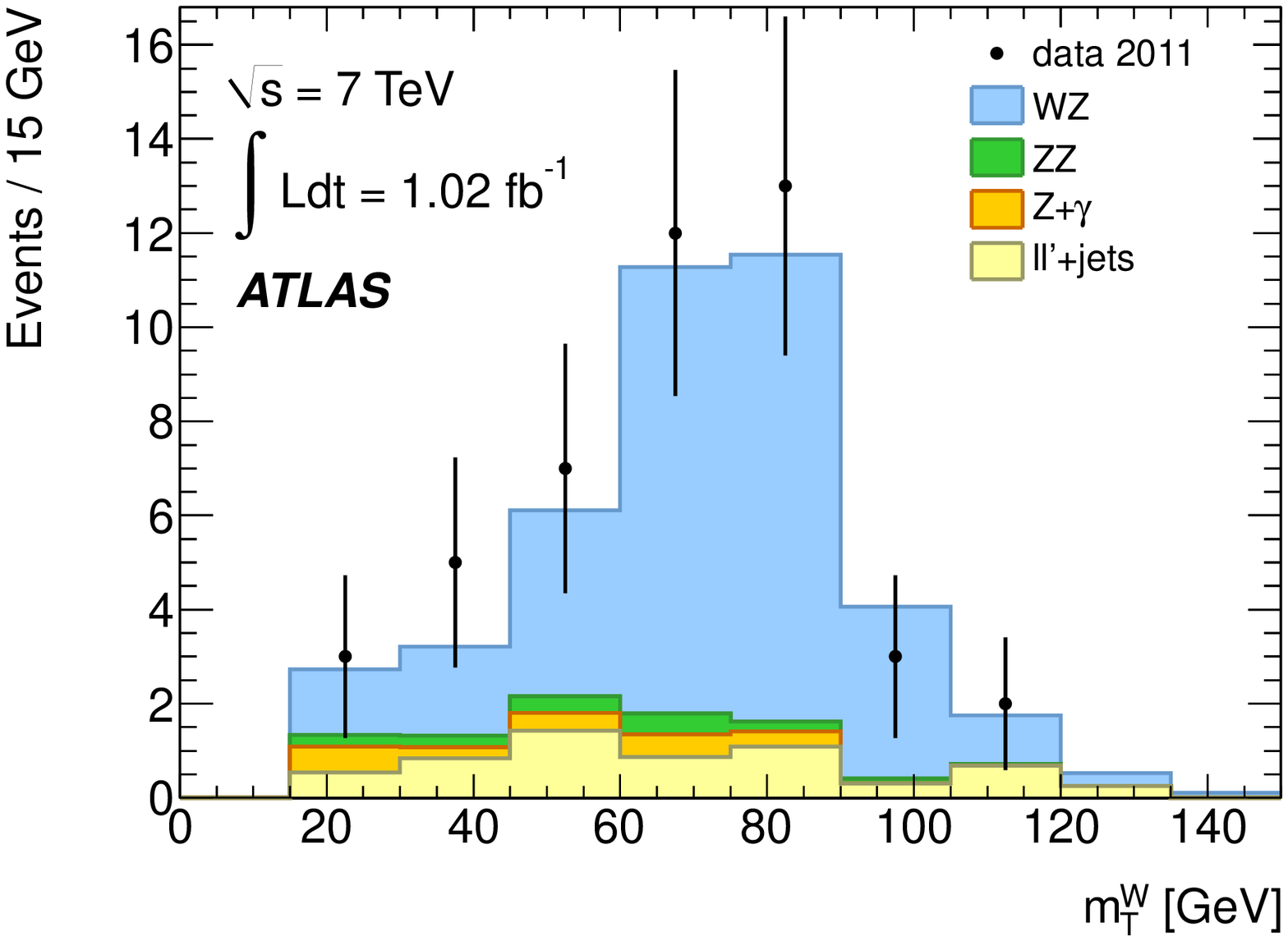}\label{f-mTW-SM}}
  \subfigure[]{\includegraphics[width=0.49\textwidth]{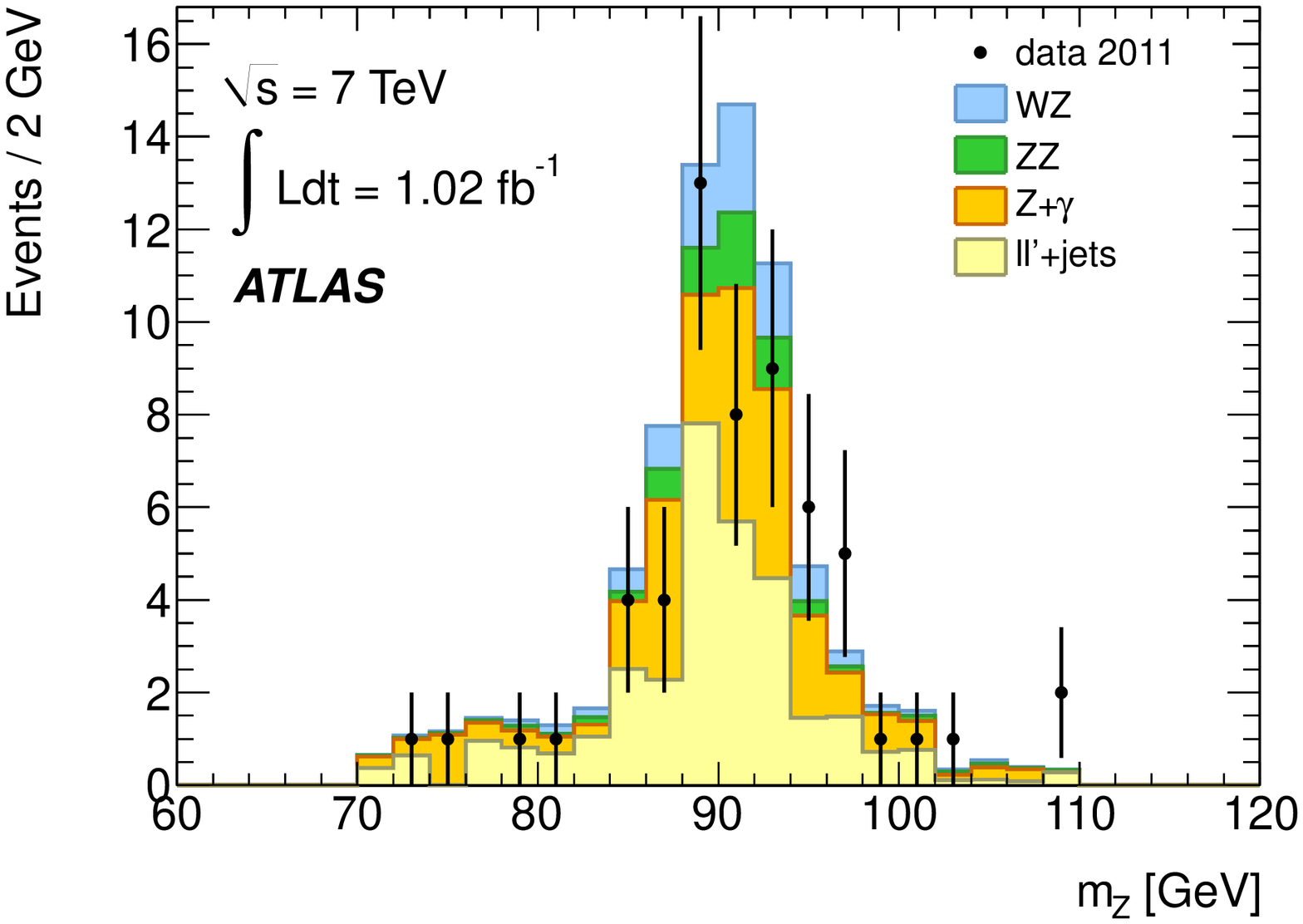}\label{f-mZ-BG}}
\caption{
    (color online) Observed and predicted $W$ boson transverse mass ($m_{\rm T}^W$) distribution in
    the SM \WZ\ control region \subref{f-mTW-SM}, and dilepton invariant mass ($m_Z$) distribution in
    the $\ell\ell^{\prime}+$jets control region \subref{f-mZ-BG}. 
 \label{fig:control}}
\end{figure*}

\section{Systematic Uncertainties}
\label{sec:systematics}
Different sources of systematic uncertainties have been considered. The first source is related to  
the lepton trigger, reconstruction and identification efficiencies. 
These efficiencies are evaluated with tag-and-probe methods using $Z
\rightarrow \ell\ell$, $W \rightarrow \ell\nu$  and $J/\psi \rightarrow \ell\ell$ events~\cite{WZinclusive}. 
Scale factors are used to correct for differences between data and MC simulation. 
The lepton trigger efficiency scale factors are compatible with unity and a systematic uncertainty of 1\% is considered.
The lepton reconstruction and identification scale factors 
are close to one 
and have a systematic uncertainty of 
1.2\% for the electrons and 0.5\% for muons~\cite{WZinclusive}. 
The lepton isolation efficiency uncertainties are estimated to be 2\% for electrons and 1\% for muons. 

The second source of uncertainty is related to the lepton energy, momentum and \met\ reconstruction.
Additional smearing is 
applied to the muon \pt\ and to the electron cluster energy in 
the simulation, so that they replicate the
$Z\rightarrow\ell\ell$ invariant mass distributions in data. The
uncertainty due to the lepton resolution smearing is of the order of 0.1\%~\cite{WZinclusive}. 
The uncertainty on the \met\ reconstruction receives contributions from 
different sources: energy deposits due to additional $pp$ collisions which are in-time and
out-of-time with respect to the bunch-crossing; energy deposits around
clusters associated to reconstructed jets and electrons; energy
deposits not associated to any reconstructed objects; and muon
momentum uncertainties. The
total systematic uncertainty on the dominant SM $WZ$ background estimation 
due to the \met\ uncertainties lies between (2$-$3)\%, depending on the channel considered. 

The third source of uncertainty is due to the limited knowledge of the theoretical 
cross sections of SM processes, used both to evaluate $WZ$, $ZZ$ and
$Z\gamma$ background contributions, and for subtracting contributions
of $W$ and $Z$ leptonic decays from the dijet sample used for the measurement of the correction factor $f$. 
An uncertainty of 7\% is assigned for the \WZ\ process, 5\% for the $ZZ$ process and 
8\% for the $Z\gamma$ process~\cite{Campbell:2011bn}, to which the MC
statistical uncertainty is added in quadrature. 

The fourth source of uncertainty is related to the uncertainty on the
$\ell \ell^{\prime}+$jets background estimation. 
The systematic uncertainty comes mainly from the uncertainty on $f$
due to differences in the kinematics and flavor composition of the QCD dijet 
events with respect to the $\ell \ell^{\prime}+$jets processes, and differences in event selection 
criteria for QCD dijet events and $WZ$ candidates. 
The factor $f$ is around 0.15 for muons and 0.07 for 
electrons over the full range of \pt\ and $\eta$, with a relative uncertainty between 5\% and 20\%. 
The estimated number of events from the $\ell\ell^{\prime}+$jets background in the
signal region using the data-driven method is $6.4 \pm 1.0 \text{(stat.)} ^{+3.2} _{-4.0} \text{(syst.)}$. 
A MC-based cross-check gives a consistent estimation of $4.3 \pm 1.1\text{(syst.)}$ events. 

The fifth source of uncertainty is related to the estimation of the signal acceptance based on MC simulation. The systematic 
uncertainty is mainly due to the choice of PDF and is found to be 0.6\% when comparing the differences between the
predictions of the nominal PDF set MRST2007 LO$^*$ and the ones given
by MSTW2008 LO \cite{Martin:2009iq}, using the standard LHAPDF
framework~\cite{Whalley:2005nh}. A cross-check has been done using the
NNPDF LO$^*$~\cite{Ball:2010de}, CT09MCS, CT09MC1 and CT09MC2~\cite{CT09MC2} PDF sets, leading to a compatible uncertainty.  

Finally the luminosity uncertainty is 3.7\%~\cite{Aad:2011dr,ATLAS-CONF-2011-116}. 

\section{Results and Interpretation}

The numbers of events expected and observed after the final selection
are reported in Table~\ref{tab:yields}. A total of 48 $WZ \to \ell\nu\ell^{\prime}\ell^{\prime}$ candidate events
are observed in data, to be compared to the SM prediction of $45.0 \pm
1.0(\rm{stat.})^{+4.6}_{-5.2}(\rm{syst.})$ events.
The expected numbers of events for a $W'$ with a mass of 750~\GeV\ and a \rhot\ with 
a mass of 500 \GeV\ are also reported.

\begin{table*}
  \caption[Data sample]{The estimated background yields, the observed number of data events, and the
predicted signal yield predicted by {\sc pythia} for a $W'$ boson with a mass of 750 \GeV\ and a \rhot\ technimeson with a mass of 500 
\GeV, are shown after applying all signal selection
cuts, for each of the four channels considered and for their combination. 
For the \rhot\ production, the relation $m_{a_{\rm T}} = 1.1 \times m_{\rho_{\rm T}}$ is used. 
Where one error is quoted, it includes all sources of systematic
uncertainty. Where two errors are given, the first comes from the limited statistics of the data and the second includes systematic uncertainties.}
\begin{center}{\small
  \begin{tabular}{l|c|c|c|c|c}
\hline
\hline
               &  $e\nu ee$       &  $\mu \nu ee$  &  $e\nu \mu\mu$  &  $\mu\nu \mu\mu$  & Combined   \\
\hline
$WZ$    & \eqparbox{col1_1}{\raggedleft 6.2} \eqparbox{col1_2}
{\raggedleft $\pm$} \eqparbox{col1_3}{\raggedleft 0.7}
\eqparbox{col1_4}{\raggedleft } \eqparbox{col1_5}{\raggedleft }
 & \eqparbox{col2_1}{\raggedleft 7.6} \eqparbox{col2_2} {\raggedleft $\pm$} \eqparbox{col2_3}{\raggedleft 0.7} \eqparbox{col2_4}{\raggedleft } \eqparbox{col2_5}{\raggedleft } & \eqparbox{col3_1}{\raggedleft 9.2} \eqparbox{col3_2} {\raggedleft $\pm$} \eqparbox{col3_3}{\raggedleft 0.8} \eqparbox{col3_4}{\raggedleft } \eqparbox{col3_5}{\raggedleft }  
             & \eqparbox{col4_1}{\raggedleft 11.6} \eqparbox{col4_2}
             {\raggedleft $\pm$} \eqparbox{col4_3}{\raggedleft 1.0} \eqparbox{col4_4}{\raggedleft } \eqparbox{col4_5}{\raggedleft }  
             & \eqparbox{col5_1}{\raggedleft 34.6} \eqparbox{col5_2}
             {\raggedleft $\pm$} \eqparbox{col5_3}{\raggedleft 3.1} \eqparbox{col5_4}{\raggedleft } \eqparbox{col5_5}{\raggedleft }   \\[1.5mm]

$ZZ$    & \eqparbox{col1_1}{\raggedleft 0.25}
\eqparbox{col1_2}{\raggedleft $^{+} _{-}$} \eqparbox{col1_3}{\raggedleft $^{0.07}_{0.11}$} \eqparbox{col1_4}{\raggedleft  } \eqparbox{col1_5}{\raggedleft } 
            & \eqparbox{col2_1}{\raggedleft 0.48}
            \eqparbox{col2_2}{\raggedleft $^{+} _{-}$}
            \eqparbox{col2_3}{\raggedleft  $^{0.14}_{0.11}$} \eqparbox{col2_4}{\raggedleft  } \eqparbox{col2_5}{\raggedleft } 
            & \eqparbox{col3_1}{\raggedleft 0.37}
            \eqparbox{col3_2}{\raggedleft $^{+} _{-}$}
            \eqparbox{col3_3}{\raggedleft  $^{0.15}_{0.11}$} \eqparbox{col3_4}{\raggedleft  } \eqparbox{col3_5}{\raggedleft } 
            & \eqparbox{col4_1}{\raggedleft 0.63}
            \eqparbox{col4_2}{\raggedleft $^{+} _{-}$}
            \eqparbox{col4_3}{\raggedleft  $^{0.16}_{0.11}$} \eqparbox{col4_4}{\raggedleft  } \eqparbox{col4_5}{\raggedleft } 
            & \eqparbox{col5_1}{\raggedleft 1.7} \eqparbox{col5_2}{\raggedleft $^{+} _{-}$} \eqparbox{col5_3}{\raggedleft  $^{0.5}_{0.3}$} \eqparbox{col5_4}{\raggedleft  }
            \eqparbox{col5_5}{\raggedleft }  \\[1.5mm]

$Z\gamma$    & \eqparbox{col1_1}{\raggedleft 1.3} \eqparbox{col1_2}
{\raggedleft $\pm$} \eqparbox{col1_3}{\raggedleft 0.7} \eqparbox{col1_4}{\raggedleft } \eqparbox{col1_5}{\raggedleft }  
                         & $-$     
                         & \eqparbox{col3_1}{\raggedleft 1.0}
                         \eqparbox{col3_2} {\raggedleft $\pm$}
                         \eqparbox{col3_3}{\raggedleft 0.9} \eqparbox{col3_4}{\raggedleft } \eqparbox{col3_5}{\raggedleft }  
                         & $-$     
                         & \eqparbox{col5_1}{\raggedleft 2.3}
                         \eqparbox{col5_2} {\raggedleft $\pm$}
                         \eqparbox{col5_3}{\raggedleft 1.1}
                         \eqparbox{col5_4}{\raggedleft }
                         \eqparbox{col5_5}{\raggedleft }
                         \\[1.5mm]

$\ell\ell^{\prime}+{\rm jets}$    & \eqparbox{col1_1}{\raggedleft 1.1} \eqparbox{col1_2}{\raggedleft $\pm$} \eqparbox{col1_3}{\raggedleft  0.4} \eqparbox{col1_4}{\raggedleft  $\pm$} \eqparbox{col1_5}{\raggedleft $0.7$} 
                                          & \eqparbox{col2_1}{\raggedleft 1.3} \eqparbox{col2_2}{\raggedleft $\pm$} \eqparbox{col2_3}{\raggedleft  0.5} \eqparbox{col2_4}{\raggedleft  $^{+} _{-}$} \eqparbox{col2_5}{\raggedleft $^{0.6}_{0.8}$} 
                                          & \eqparbox{col3_1}{\raggedleft 3.0} \eqparbox{col3_2}{\raggedleft $\pm$} \eqparbox{col3_3}{\raggedleft  0.7} \eqparbox{col3_4}{\raggedleft  $^{+} _{-}$} \eqparbox{col3_5}{\raggedleft $^{1.6}_{1.9}$} 
                                          & \eqparbox{col4_1}{\raggedleft 1.0} \eqparbox{col4_2}{\raggedleft $\pm$} \eqparbox{col4_3}{\raggedleft  0.4} \eqparbox{col4_4}{\raggedleft  $^{+} _{-}$} \eqparbox{col4_5}{\raggedleft $^{0.5}_{0.6}$} 
                                          & \eqparbox{col5_1}{\raggedleft 6.4} \eqparbox{col5_2}{\raggedleft $\pm$} \eqparbox{col5_3}{\raggedleft  1.0} \eqparbox{col5_4}{\raggedleft  $^{+} _{-}$} \eqparbox{col5_5}{\raggedleft $^{3.2}_{4.0}$}  \\[1.5mm]

Overall backgrounds                     & \eqparbox{col1_1}{\raggedleft 8.9} \eqparbox{col1_2}{\raggedleft $\pm$} \eqparbox{col1_3}{\raggedleft  0.4} \eqparbox{col1_4}{\raggedleft  $\pm$} \eqparbox{col1_5}{\raggedleft $1.2$} 
                                          & \eqparbox{col2_1}{\raggedleft 9.4} \eqparbox{col2_2}{\raggedleft $\pm$} \eqparbox{col2_3}{\raggedleft  0.5} \eqparbox{col2_4}{\raggedleft  $^{+} _{-}$} \eqparbox{col2_5}{\raggedleft $^{0.9}_{1.1}$} 
                                          & \eqparbox{col3_1}{\raggedleft 13.6} \eqparbox{col3_2}{\raggedleft $\pm$} \eqparbox{col3_3}{\raggedleft  0.7} \eqparbox{col3_4}{\raggedleft  $^{+} _{-}$} \eqparbox{col3_5}{\raggedleft $^{2.0}_{2.3}$} 
                                          & \eqparbox{col4_1}{\raggedleft 13.2} \eqparbox{col4_2}{\raggedleft $\pm$} \eqparbox{col4_3}{\raggedleft  0.4} \eqparbox{col4_4}{\raggedleft  $^{+} _{-}$} \eqparbox{col4_5}{\raggedleft $^{1.3}_{1.2}$} 
                                          & \eqparbox{col5_1}{\raggedleft 45.0} \eqparbox{col5_2}{\raggedleft $\pm$} \eqparbox{col5_3}{\raggedleft  1.0} \eqparbox{col5_4}{\raggedleft  $^{+} _{-}$} \eqparbox{col5_5}{\raggedleft $^{4.6}_{5.2}$}  \\[1.5mm]
Data             & 9 & 7 & 16 & 16 & 48 \\ [1.5mm]
\hline

$W' \rightarrow WZ$ ($m_{W'}=750$~GeV)	&
\eqparbox{col1_1}{\raggedleft 0.74} \eqparbox{col1_2} {\raggedleft
  $\pm$} \eqparbox{col1_3}{\raggedleft 0.07} \eqparbox{col1_4}{\raggedleft } \eqparbox{col1_5}{\raggedleft }  
& \eqparbox{col2_1}{\raggedleft 0.82} \eqparbox{col2_2} {\raggedleft
  $\pm$} \eqparbox{col2_3}{\raggedleft 0.06} \eqparbox{col2_4}{\raggedleft } \eqparbox{col2_5}{\raggedleft }  
& \eqparbox{col3_1}{\raggedleft 0.97} \eqparbox{col3_2} {\raggedleft
  $\pm$} \eqparbox{col3_3}{\raggedleft 0.06} \eqparbox{col3_4}{\raggedleft } \eqparbox{col3_5}{\raggedleft }  
& \eqparbox{col4_1}{\raggedleft 1.10} \eqparbox{col4_2} {\raggedleft
  $\pm$} \eqparbox{col4_3}{\raggedleft 0.08} \eqparbox{col4_4}{\raggedleft } \eqparbox{col4_5}{\raggedleft }  
& \eqparbox{col5_1}{\raggedleft 3.64}  \eqparbox{col5_2} {\raggedleft
  $\pm$} \eqparbox{col5_3}{\raggedleft  0.21}
\eqparbox{col5_4}{\raggedleft } \eqparbox{col5_5}{\raggedleft }
\\[1.5mm]

$\rho_{\rm T} \rightarrow WZ$ ($m_{\rm \rho_{\rm T}}=500$~GeV)	&
\eqparbox{col1_1}{\raggedleft 0.68} \eqparbox{col1_2} {\raggedleft
  $\pm$} \eqparbox{col1_3}{\raggedleft 0.08} \eqparbox{col1_4}{\raggedleft } \eqparbox{col1_5}{\raggedleft }  
& \eqparbox{col2_1}{\raggedleft  0.79} \eqparbox{col2_2} {\raggedleft  $\pm$} \eqparbox{col2_3}{\raggedleft  0.08} \eqparbox{col2_4}{\raggedleft } \eqparbox{col2_5}{\raggedleft }  
& \eqparbox{col3_1}{\raggedleft  0.97} \eqparbox{col3_2} {\raggedleft
  $\pm$} \eqparbox{col3_3}{\raggedleft  0.09} \eqparbox{col3_4}{\raggedleft } \eqparbox{col3_5}{\raggedleft }  
& \eqparbox{col4_1}{\raggedleft 1.11} \eqparbox{col4_2} {\raggedleft $\pm$} \eqparbox{col4_3}{\raggedleft 0.10} \eqparbox{col4_4}{\raggedleft } \eqparbox{col4_5}{\raggedleft }  
& \eqparbox{col5_1}{\raggedleft 3.55} \eqparbox{col5_2} {\raggedleft $\pm$} \eqparbox{col5_3}{\raggedleft 0.24} \eqparbox{col5_4}{\raggedleft } \eqparbox{col5_5}{\raggedleft }   \\[1.5mm]
\hline
\hline
  \end{tabular}}
    \label{tab:yields}
\end{center}
\end{table*}

The overall acceptance times trigger, reconstruction and selection
efficiencies ($A \times \epsilon$) for EGM $W' \to WZ \to \ell\nu\ell^{\prime}\ell^{\prime}$ and
the LSTC $\rhot \to WZ \to \ell\nu\ell^{\prime}\ell^{\prime}$ events as implemented in
\pythia\ is shown in Table~\ref{tab:acceptance} for various $WZ$ resonance masses. 
The value of $A \times \epsilon$ is 6.2\% for $m_{W'}=200$ GeV and increases to 
20.5\% for $m_{W'}=1$ TeV.  
The corresponding $A \times \epsilon$ for the LSTC \rhot~is found to be slightly lower 
than that of the EGM $W'$ due to the fact that the \pythia\ implementation of the $\rhot
\to W Z$ process does not account for the polarizations of vector bosons in their decay. 
A massive $W'$ boson is expected to decay predominantly to longitudinally polarized 
$W$ and $Z$ bosons, as is the \rhot\ technimeson.
While the production and decay with spin correlations is fully 
implemented in \pythia\ for $W'$, spin correlation information is not considered in the
decay of the $W$ and $Z$ bosons in the \rhot\ case, hence they each decay isotropically 
in their respective rest frames. 
This leads to a softer lepton \pt\ spectrum and consequently lower $A \times \epsilon$. 
The interpretation of the data in terms of \rhot\ production is performed in two different manners:
the first uses the \pythia\ implementation of \rhot\ production and
decay, and 
the second assumes that $A \times \epsilon$ for the \rhot\ is equal to that of the $W'$. 

\begin{table}[h!]
  \caption{Signal $A \times \epsilon$ for $W' \to WZ \to
    \ell\nu\ell'\ell'$ and $\rhot \to WZ \to \ell\nu\ell'\ell'$ samples as
    implemented in \pythia, with statistical uncertainties. Missing values for $\rho_{\rm T}$
    correspond to signal samples not considered.}
  \begin{center}{
      \begin{tabular}{c|cc}
        \hline
        \hline
        Mass [\GeV] & $A \times \epsilon$ for $W'$ (\%) & ~~~~$A \times \epsilon$ for \rhot\ (\%) \\
        \hline
        200  &  $\ \ 6.2 \pm 0.2$   &   $\ \ 5.7 \pm 0.2$ \\
        250  &  $\ \ 8.2 \pm 0.4$   &   $\ \ 6.1 \pm 0.2$ \\
        300  &  $10.0 \pm 0.5$  &   $\ \ 7.6 \pm 0.3$ \\
        350  &  $11.6 \pm 0.3$  &   $\ \ 9.4 \pm 0.3$ \\
        400  &  $13.2 \pm 0.5$  &   $10.8 \pm 0.3$ \\
        450  &  $14.5 \pm 0.6$  &   $11.8 \pm 0.3$ \\
        500  &  $15.9 \pm 0.3$  &   $12.6 \pm 0.3$ \\
        550  &  $16.9 \pm 0.6$  &   -- \\ 
        600  &  $17.9 \pm 0.6$  &   $13.8 \pm 0.3$ \\ 
        650  &  $18.7 \pm 0.6$  &   -- \\
        700  &  $19.4 \pm 0.7$  &   $15.6 \pm 0.4$ \\
        750  &  $19.9 \pm 0.3$  &   -- \\
        800  &  $20.3 \pm 0.7$  &   $16.1 \pm 0.4$ \\
        850  &  $20.6 \pm 0.7$  &   -- \\
        900  &  $20.6 \pm 0.7$  &   -- \\
        950  &  $20.6 \pm 0.7$  &   -- \\
        1000 &  $20.5 \pm 0.3$  &   -- \\
        \hline
        \hline
      \end{tabular}
    \label{tab:acceptance}}
  \end{center}
\end{table}
      
The transverse mass distribution of the $WZ$ candidates is presented in Fig.~\ref{fig:wz} 
for data and background expectations together with possible contributions from $W'$ and $\rho_T$ using {\sc pythia}.  
The $\ell\ell^{\prime}+$jets and $Z\gamma$ background contributions to the \mtwz\ distribution are extrapolated using exponential functions to extend over
the full \mtwz\ signal region.
The transverse mass distribution is used to build a log-likelihood ratio (LLR) test statistic~\cite{kendall}, which allows 
the compatibility of the data with the presence of a signal in addition to the background to be assessed, 
in a modified frequentist approach~\cite{Fisher:2006zz}. 
Confidence levels for the signal plus background
hypothesis, $\rm{CL}_{s+b}$, and background-only hypothesis, $\rm{CL}_b$, are computed by
integrating the LLR distributions obtained from simulated
pseudo-experiments using Poisson statistics. The confidence
level for the signal hypothesis $\rm{CL}_{s}$, defined as the ratio
$\rm{CL}_{s+b}/\rm{CL}_b$, is used to determine the exclusion limits. 

\begin{figure}[b!]
  \centering
  \includegraphics[width=0.5\textwidth]{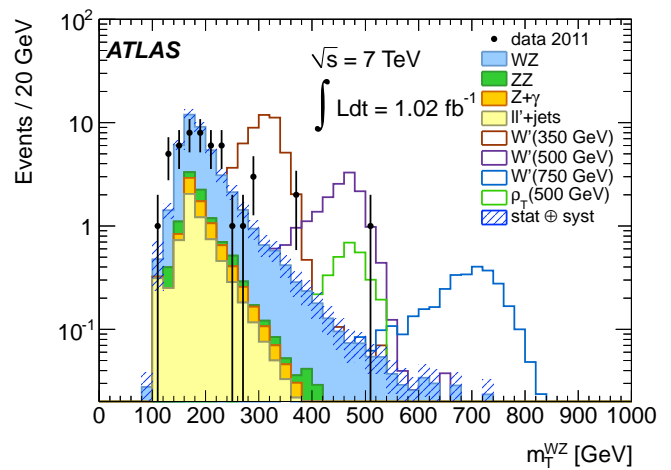} 
  \caption{(color online) Observed and predicted \mtwz\ distribution for events with all selection cuts applied. 
    Predictions from three $W'$ samples with masses of 350 GeV, 500 GeV and 750 GeV and a \rhot\ sample with 
    a mass of 500 GeV using {\sc pythia} are also shown.
  \label{fig:wz}}
\end{figure}

The probability that the background fluctuations give rise to an excess at least as large as that observed in data has been 
computed as $p$-value $=1-\rm{CL}_b$ and is reported in Table~\ref{tab:limitW} 
for the signal hypothesis of a $W'$ particle with mass from 200 GeV to 1 TeV. 
Since no statistically significant excess is observed for any value of
the $W'$ mass, limits are derived on the production cross section
times branching ratio ($\sigma \times BR(W' \to WZ)$) for a $W'$ decaying to
$WZ$, already corrected for the $A \times \epsilon$ of the leptonic decay $WZ \to \ell \nu \ell^{\prime} \ell^{\prime}$. 
The 95\% CL limit on $\sigma \times BR(W' \to WZ)$ is defined as the 
value giving $\rm{CL}_{s} = 0.05$. 
The upper limit on $\sigma \times BR(W' \to WZ)$ 
for $pp \rightarrow W' \rightarrow WZ$ 
as a function of the $W'$ mass is shown in Fig.~\ref{fig:1Dlimits}(a) and the values are reported in Table~\ref{tab:limitW}. 
Simulation of $W'$ bosons is performed for $m_{W'}$ between 200 GeV and 1 TeV with a 150 to 250 GeV mass spacing, 
and an interpolation procedure provides $m_T^{WZ}$ shape templates with a 50 GeV spacing.
The \mtwz\ shapes from the 
fully simulated signal samples have been fitted with a Crystal Ball function 
using RooFit~\cite{RooFit}. The obtained
Crystal Ball parameters are fitted as a function of the
$W'$ mass and the functional value for these parameters is then used
to build the \mtwz\ templates for the intermediate $W'$ mass points.
The observed (expected) exclusion limit on the $W'$ mass is found to be $760~(776)$~\GeV. 

\begin{table}[h!]
  \caption{Expected and observed limit on the $\sigma \times BR(W' \to WZ)$
    [pb] for $W'$ production decaying to $WZ$, as a
    function of the $W'$ mass. The $p$-values are also reported.} 
  \begin{center}{
      \begin{tabular}{c|cc|c}
        \hline
        \hline
        $W'$ Mass [\GeV]& \multicolumn{2}{c|}{Excluded $\sigma \times BR(W' \to WZ)$ [pb]} & $p$-value \\ 
        \cline{2-3}
        & Expected & Observed \\
        \hline
        200  &  7.31    &    7.62 &  0.43\\ 
        250  &  5.26    &    6.55 &  0.34\\ 
        300  &  2.74    &    3.38 &  0.28\\ 
        350  &  1.72    &    2.06 &  0.25\\ 
        400  &  1.18    &    1.48 &  0.25\\ 
        450  &  0.92   &     1.07 &  0.23\\ 
        500  &  0.76   &     0.93 &  0.21\\ 
        550  &  0.61   &     0.79 &  0.19\\ 
        600  &  0.54   &     0.63 &  0.26\\ 
        650  &  0.51   &     0.56 &  0.33\\ 
        700  &  0.48   &     0.53 &  0.34\\ 
        750  &  0.49   &     0.52 &  0.34\\ 
        800  &  0.45   &     0.50 &  0.37\\ 
        850  &  0.46   &     0.47 &  0.38\\ 
        900  &  0.50   &     0.50 &  0.39\\ 
        950  &  0.44   &     0.44 &  0.40\\ 
        1000 &   0.48 &     0.46  & 0.35\\ 
        \hline
        \hline
      \end{tabular}
    \label{tab:limitW}}
  \end{center}
\end{table}

\begin{figure*}
  \centering
  \subfigure[]{\includegraphics[width=0.45\textwidth]{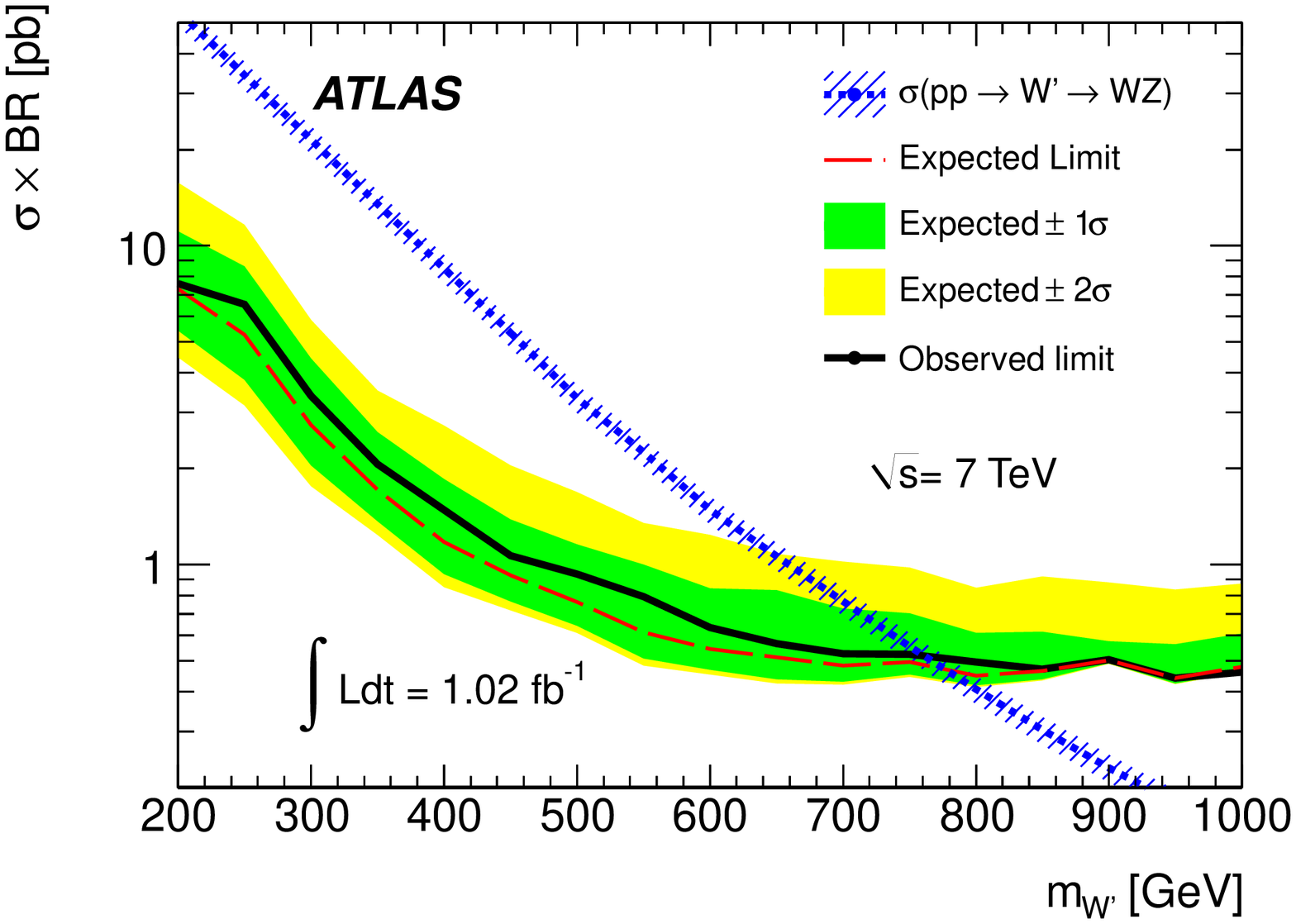}\label{f-wprime-limit}}
  \subfigure[]{\includegraphics[width=0.45\textwidth]{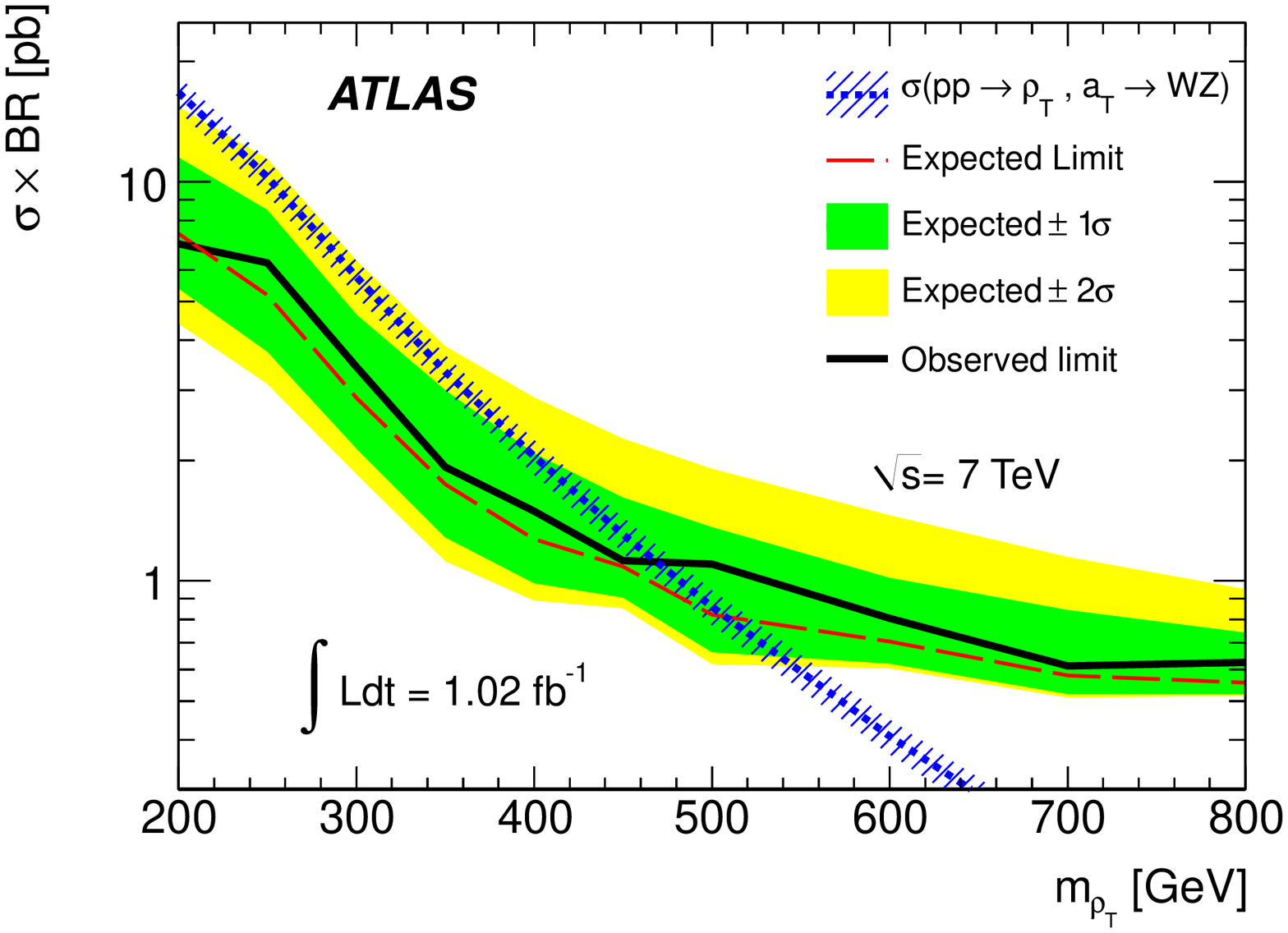}\label{f-technirho-limit}}
\caption{
    The observed and expected limits on $\sigma \times BR(W' \to WZ)$ 
    for $W' \rightarrow WZ$ \subref{f-wprime-limit} and $pp \rightarrow \rho_T, a_T \rightarrow WZ$ \subref{f-technirho-limit}. 
    The theoretical prediction is shown with a systematic uncertainty of 5\% due to the choice of PDF and is estimated by 
comparing the differences between the
predictions of the nominal PDF set MRST2007 LO$^*$ and the ones given
by MSTW2008 LO PDF using the LHAPDF framework.
The green and yellow bands represent respectively the $1\sigma$ and $2\sigma$ uncertainty on the expected limit.
  \label{fig:1Dlimits}}
\end{figure*}

The observed (expected) limits on $\sigma \times BR(\rhot\ \to WZ)$ 
for the \rhot\ technimeson are presented
in Fig.~\ref{fig:1Dlimits}(b) assuming $\Ma = 1.1 \Mrho$ and unpolarized $W$ and $Z$ decays. 
This corresponds to an observed (expected) limit on the \rhot\ mass of $467~(506)$~\GeV.
A limit on the \rhot\ mass of $456~(482)$~\GeV\ is obtained if $\Ma \gg \Mrho$. 
Assuming $A \times \epsilon$ for the \rhot\ signal to be equal to that of the $W'$ signal, 
which is estimated by accounting for predominantly longitudinal $W$ and $Z$ polarization, 
the observed (expected) limit on the \rhot\ mass is $483~(553)$~\GeV\ for $\Ma = 1.1 \Mrho$, and 
$469~(507)$~\GeV\ for $\Ma \gg \Mrho$. 
Table~\ref{tab:limitrhot_summary} summarizes these limits, which all assume the relation $m_{\rho_T} = m_{\pi_T} + m_W$. 

\begin{table}[h!]
  \caption{Observed (expected) limit on the \rhot\ mass with two different 
   assumptions about $A \times \epsilon$ for \rhot\ and two mass hierarchy 
   assumptions between $a_{\rm T}$ and \rhot.}
  \begin{center}{
      \begin{tabular}{c|cc}
        \hline
        \hline
         & \multicolumn{2}{c}{Excluded \rhot\ mass [GeV] }\\
         & $\Ma = 1.1 \Mrho$ & ~~~~~~$\Ma \gg \Mrho$ \\
        \hline
        $A \times \epsilon$ from $W'$ sample  & 483 (553) & ~~~~~~ 469 (507) \\ [1.5mm]
        $A \times \epsilon$ from \rhot\ sample  & 467 (506)  & ~~~~~~ 456 (482)  \\
        \hline
        \hline
      \end{tabular}
    \label{tab:limitrhot_summary}}
  \end{center}
\end{table}

Figure~\ref{rho_pi_exclusion}
shows the 95\% CL expected and observed excluded regions 
in the ($m_{\rho_T}$, $m_{\pi_T}$) plane for $\Ma = 1.1 \Mrho$ and $\Ma \gg \Mrho$, respectively. 
Results are shown under the two assumptions on $A \times \epsilon$ for the \rhot\ signal.

\begin{figure*}
  \centering
  \subfigure[]{\includegraphics[width=0.45\textwidth]{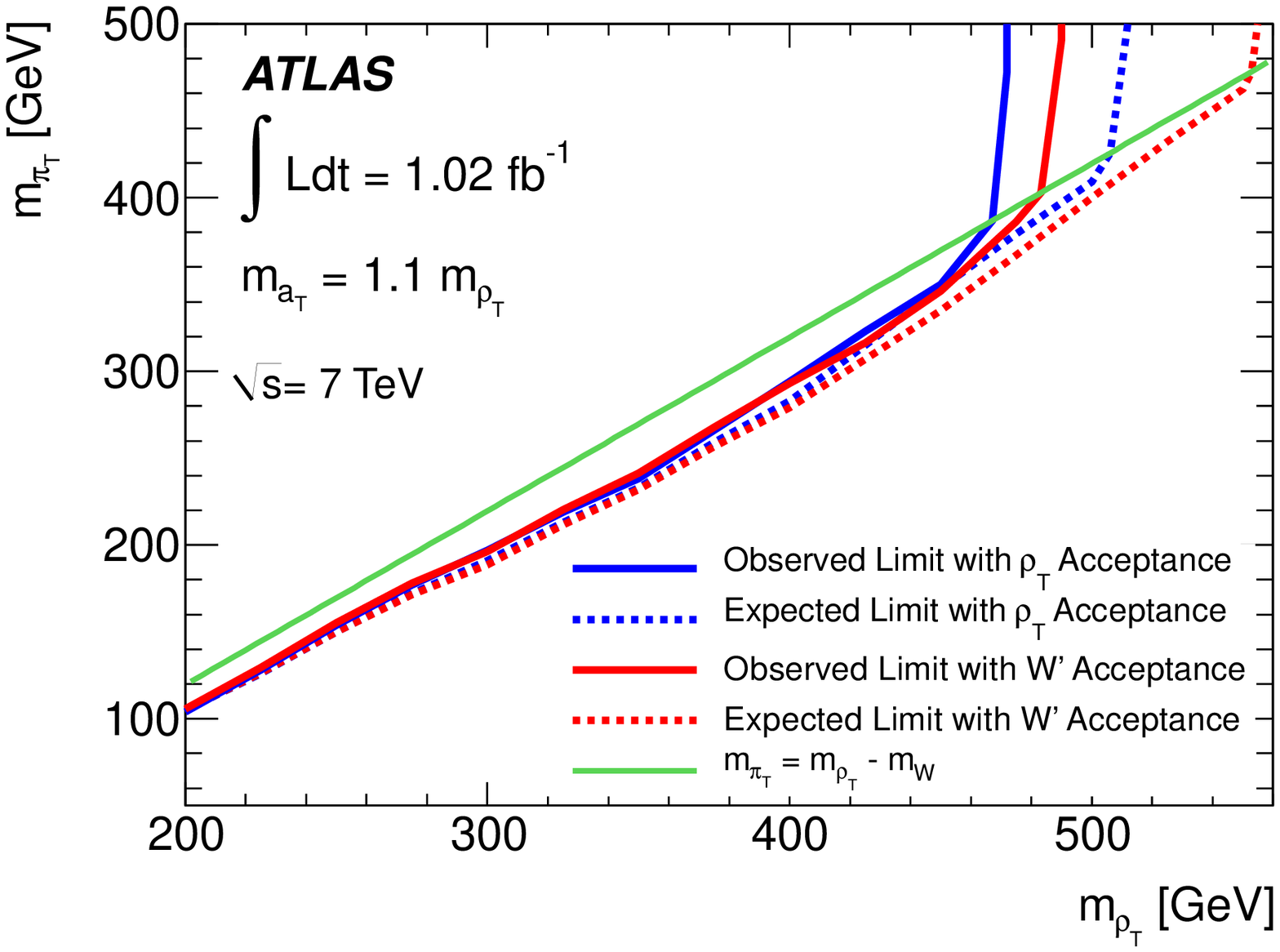}\label{f-Rho-vs-Pi-withaT}}
  \subfigure[]{\includegraphics[width=0.45\textwidth]{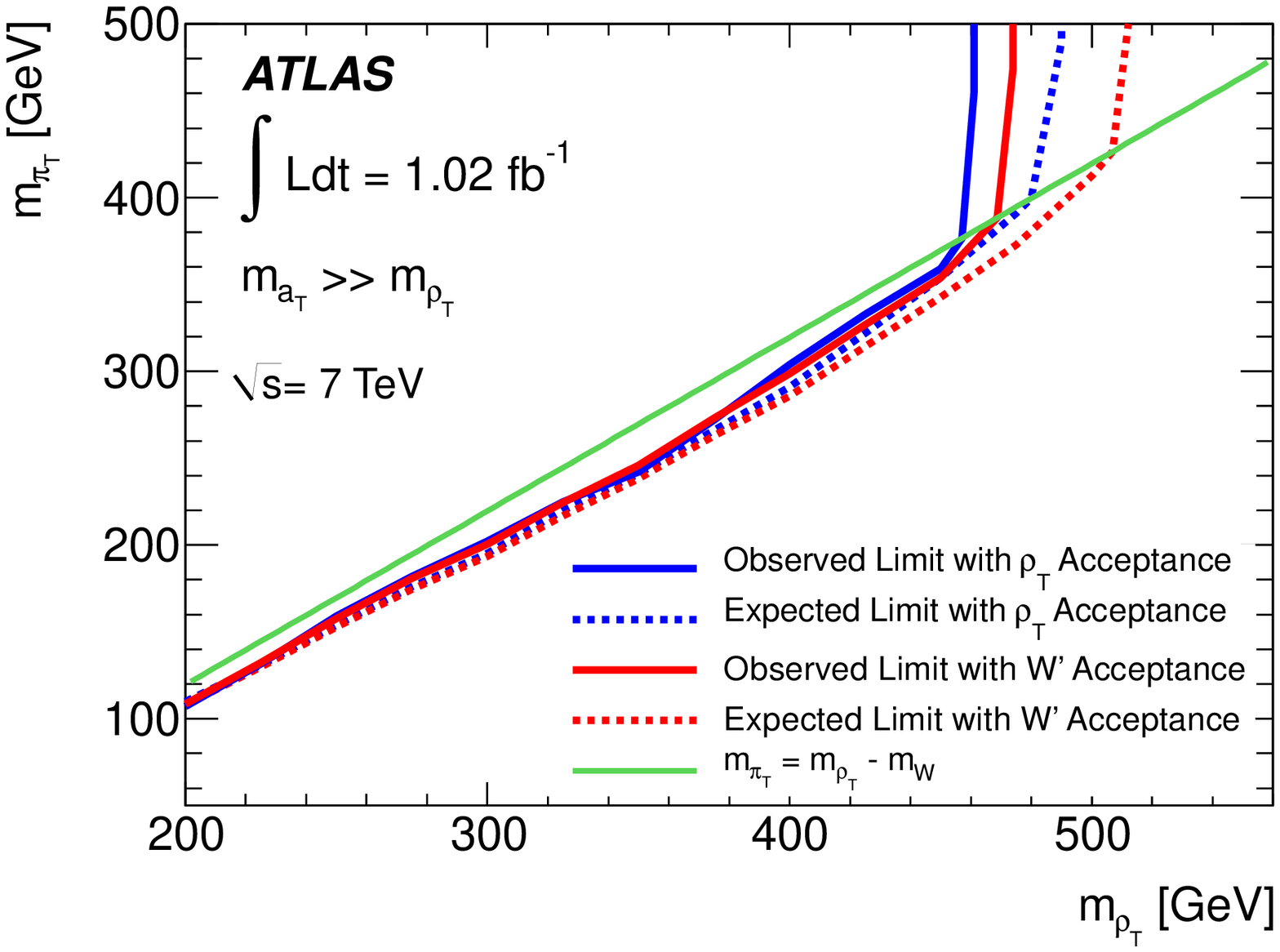}\label{f-Rho-vs-Pi-noaT}}
  \caption{The $95$\% CL expected and observed excluded mass regions in the 
           (\Mrho, $m_{\pi_{\rm T}}$) plane for $\Ma = 1.1 \Mrho$
           \subref{f-Rho-vs-Pi-withaT} and $\Ma \gg \Mrho$
           \subref{f-Rho-vs-Pi-noaT}, above the curves. 
           Two different assumptions about the \rhot\ signal $A \times \epsilon$ are used: assuming a \rhot\ signal 
           where $A \times \epsilon$ is equal to that of the $W'$ signal and assuming a \rhot\ signal 
           where $A \times \epsilon$ is obtained through its implementation in {\sc pythia}. 
 \label{rho_pi_exclusion}}
\end{figure*}

\section{Conclusion} 
\label{sec:conclusion}
A search for resonant production of a pair of \WZ\ bosons with
three charged leptons in the final state has been performed using 1.02 fb$^{-1}$ of data 
collected with the ATLAS detector in $pp$ collisions at $\sqrt{s}=7$ \TeV~at the Large Hadron Collider.
No significant excess of events is observed and
upper limits are derived on the production cross section times branching ratio of new physics using 
the transverse mass of the $WZ$ system.
EGM $W'$ bosons with masses up to $760$~\GeV\ are excluded at 95\% CL. 
Using the mass hierarchy assumption $m_{\rho_{\rm T}} = m_{\pi_{\rm
    T}} + m_W$, LSTC \rhot\ technimesons with masses from $200$ \GeV\ up to $467$~\GeV\ and $456$~\GeV\ are excluded at 95\% CL 
for $\Ma = 1.1 \Mrho$ and $\Ma \gg \Mrho$ respectively using the \pythia\ implementation of 
\rhot\ production. Assuming the kinematics of the $W'$ production and decay are valid for 
the $\rho_{\rm T}$ technimeson, \rhot\ with masses from $200$ \GeV\ up to  $483$~\GeV\ and $469$~\GeV\ are 
excluded for $\Ma = 1.1 \Mrho$ and $\Ma \gg \Mrho$ respectively.

\section{Acknowledgements}
We thank CERN for the very successful operation of the LHC, as well as the
support staff from our institutions without whom ATLAS could not be
operated efficiently.

We acknowledge the support of ANPCyT, Argentina; YerPhI, Armenia; ARC,
Australia; BMWF, Austria; ANAS, Azerbaijan; SSTC, Belarus; CNPq and FAPESP,
Brazil; NSERC, NRC and CFI, Canada; CERN; CONICYT, Chile; CAS, MOST and NSFC,
China; COLCIENCIAS, Colombia; MSMT CR, MPO CR and VSC CR, Czech Republic;
DNRF, DNSRC and Lundbeck Foundation, Denmark; ARTEMIS and ERC, European Union;
IN2P3-CNRS, CEA-DSM/IRFU, France; GNAS, Georgia; BMBF, DFG, HGF, MPG and AvH
Foundation, Germany; GSRT, Greece; ISF, MINERVA, GIF, DIP and Benoziyo Center,
Israel; INFN, Italy; MEXT and JSPS, Japan; CNRST, Morocco; FOM and NWO,
Netherlands; RCN, Norway; MNiSW, Poland; GRICES and FCT, Portugal; MERYS
(MECTS), Romania; MES of Russia and ROSATOM, Russian Federation; JINR; MSTD,
Serbia; MSSR, Slovakia; ARRS and MVZT, Slovenia; DST/NRF, South Africa;
MICINN, Spain; SRC and Wallenberg Foundation, Sweden; SER, SNSF and Cantons of
Bern and Geneva, Switzerland; NSC, Taiwan; TAEK, Turkey; STFC, the Royal
Society and Leverhulme Trust, United Kingdom; DOE and NSF, United States of
America.

The crucial computing support from all WLCG partners is acknowledged
gratefully, in particular from CERN and the ATLAS Tier-1 facilities at
TRIUMF (Canada), NDGF (Denmark, Norway, Sweden), CC-IN2P3 (France),
KIT/GridKA (Germany), INFN-CNAF (Italy), NL-T1 (Netherlands), PIC (Spain),
ASGC (Taiwan), RAL (UK) and BNL (USA) and in the Tier-2 facilities
worldwide.

\bibliographystyle{apsrev4-1}
\bibliography{WZResPaper}
\onecolumngrid
\clearpage
\input{atlas_authlist.tex}

\end{document}

%% file: atlas_authlist.tex
\begin{flushleft}
{\Large The ATLAS Collaboration}

\bigskip

G.~Aad$^{\rm 48}$,
B.~Abbott$^{\rm 112}$,
J.~Abdallah$^{\rm 11}$,
S.~Abdel~Khalek$^{\rm 116}$,
A.A.~Abdelalim$^{\rm 49}$,
A.~Abdesselam$^{\rm 119}$,
O.~Abdinov$^{\rm 10}$,
B.~Abi$^{\rm 113}$,
M.~Abolins$^{\rm 89}$,
O.S.~AbouZeid$^{\rm 159}$
% ...
\bigskip

% ...
$^{10}$ Institute of Physics, Azerbaijan Academy of Sciences, Baku, Azerbaijan\\
$^{11}$ Institut de F\'isica d'Altes Energies and Departament de F\'isica de la Universitat Aut\`onoma  de Barcelona and ICREA, Barcelona, Spain\\
% ...
$^{48}$ Fakult\"{a}t f\"{u}r Mathematik und Physik, Albert-Ludwigs-Universit\"{a}t, Freiburg i.Br., Germany\\
$^{49}$ Section de Physique, Universit\'e de Gen\`eve, Geneva, Switzerland\\
% ...
$^{89}$ Department of Physics and Astronomy, Michigan State University, East Lansing MI, United States of America\\
% ...
$^{112}$ Homer L. Dodge Department of Physics and Astronomy, University of Oklahoma, Norman OK, United States of America\\
$^{113}$ Department of Physics, Oklahoma State University, Stillwater OK, United States of America\\
% ...
$^{116}$ LAL, Univ. Paris-Sud and CNRS/IN2P3, Orsay, France\\
% ...
$^{119}$ Department of Physics, Oxford University, Oxford, United Kingdom\\
% ...
$^{159}$ Department of Physics, University of Toronto, Toronto ON, Canada\\
% ...
\end{flushleft}


%% file: WZResPaper.bbl
\begin{thebibliography}{42}
\bibitem [{geo()}]{geometry}%
  \BibitemOpen
  \href@noop {} {}\bibinfo {note} {ATLAS uses a right-handed coordinate system
  with its origin at the nominal interaction point~(IP) in the center of the
  detector and the $z$-axis along the beam pipe. The $x$-axis points from the
  IP to the center of the LHC ring, and the $y$ axis points upward. Cylindrical
  coordinates $(R,\phi)$ are used in the transverse plane, $\phi$ being the
  azimuthal angle around the beam pipe. The pseudorapidity is defined in terms
  of the polar angle $\theta$ as $\eta=-\ln\tan(\theta/2)$. The transverse
  energy $E_{\rm T}$ is defined as $E \sin{\theta}$, where $E$ is the energy
  associated to the calorimeter cell or energy cluster. Similarly, \pt\ is the
  momentum component transverse to the beam line.}\BibitemShut {Stop}%
\end{thebibliography}
